\documentclass[aps,prl,reprint,superscriptaddress,amsmath,amssymb,floatfix,showpacs]{revtex4-1}
\usepackage{mathtools}
\usepackage{amsmath}
\usepackage{graphicx}
\usepackage{esint}
\usepackage{times}

\makeatletter
\usepackage{color}
\usepackage{epstopdf}
\usepackage{slashed}
\usepackage{bm}
\usepackage{braket}
\usepackage{soul}

\makeatother
\begin{document}

\title{
Phase diagram and
quantum criticality of Heisenberg spin chains with Ising-like interchain couplings -- Implication to YbAlO$_3$
}

\author{Yuchen Fan}
\affiliation{Department of Physics and Beijing Key Laboratory of Opto-electronic Functional Materials and Micro-nano Devices, Renmin University of China, Beijing 100872, China
}

\author{Jiahao Yang}
\affiliation{Tsung-Dao Lee Institute \& School of Physics and Astronomy, Shanghai Jiao Tong University, Shanghai 200240, China}

\author{Weiqiang Yu}
\affiliation{Department of Physics and Beijing Key Laboratory of Opto-electronic Functional Materials and Micro-nano Devices, Renmin University of China, Beijing 100872, China
}

\author{Jianda Wu}
\affiliation{Tsung-Dao Lee Institute \& School of Physics and Astronomy, Shanghai Jiao Tong University, Shanghai 200240, China}

\author{Rong Yu}
\affiliation{Department of Physics and Beijing Key Laboratory of Opto-electronic Functional Materials and Micro-nano Devices, Renmin University of China, Beijing 100872, China
}

\begin{abstract}
Motivated by recent progress on field-induced phase transitions in quasi-one-dimensional quantum antiferromagnets, we study the phase diagram of 
$S=1/2$ antiferromagnetic
Heisenberg chains with Ising anisotropic interchain couplings under
a longitudinal magnetic field 
via
large-scale quantum Monte Carlo simulations.
The interchain interactions is shown to
enhance longitudinal spin correlations to stabilize an
incommensurate longitudinal
spin density wave order 
at low temperatures.
With increasing field
the ground state changes to a canted antiferromagnetic order
until the magnetization fully saturates above a quantum critical
point controlled by the $(3+2)$D XY universality.
Increasing temperature in the quantum
critical regime 
the
system experiences
a fascinating dimension crossover
to 
a
universal Tomonaga-Luttinger liquid. 
The calculated NMR relaxation rate $1/T_1$ indicates
this Luttinger liquid behavior 
survives 
a broad field and temperature regime. 
Our results determine the global phase diagram and quantitative features of quantum criticality of a general model for 
quasi-one-dimensional spin chain compounds, 
and thus lay down a concrete ground to the study on these materials.
\end{abstract}
\maketitle

{\it Introduction.~} In low-dimensional correlated electron systems strong quantum fluctuations give rise to quantum phase transitions (QPTs)~\cite{Sachdev_book:2011} and a number of exotic quantum phenomena, such as unconventional superconductivity~\cite{Lee_RMP:2006,Cao_Nat:2018}, non-Fermi liquid behavior~\cite{Stewart_RMP:2001,Loehneysen_JLTP:2010}, and quantum spin liquids~\cite{Zheng_PRL:2017}. In the past decade, tremendous progresses have been made in understanding the nature of QPTs and associated emerging phenomena in quasi-one-dimensional (Q1D) antiferromagnets. These include the $E_8$ symmetry~\cite{Zam_E8:1989, Coldea_Sci:2010, Wu_E8:2014}, many-body string excitations~\cite{Wang_Nat:2018,Wang_PRL:2019} and novel quantun criticality~\cite{Faure_NP:2018,Cui_PRL:2019} in transverse field Ising chains, and Bose-Einstein condensation (BEC) and glassy phases in coupled antiferromagnetic (AFM) chains~\cite{Zapf_RMP:2014,Yu_Nat:2012}.

As a paradigmatic model for 1D quantum antiferromagnets, the
$S=1/2$
Heisenberg chain is well described by a Tomonaga-Luttinger liquid (TLL),
where both the longitudinal and transverse spin correlation functions follow algebraic decay.\cite{Giamarchi_book:2004} Under a magnetic field, the staggered transverse correlations are always dominant over the longitudinal ones, and a canted AFM order with staggered transverse correlations (denoted as the TAF order) is stabilized when interchain couplings become relevant. In systems with an Ising anisotropy, besides the TAF phase which arises from a spin-flop mechanism~\cite{Fisher_PRL:1974}, the peculiar quantum fluctuations in the Ising anisotropic XXZ chain give rise to incommensurate modulation of the longitudinal spin correlations~\cite{Haldane_PRL:1980} and can stabilize an incommensurate longitudinal spin density wave (LSDW) order~\cite{Okunishi_PRB:2007}. This LSDW state has been recently observed in several Q1D antiferromagnets~\cite{Kimura_PRL:2008,Itoh_PRL:2008,Grenier_PRB:2015,Klanjsek_PRB:2015,Shen_NJP:2019}.

Recent inelastic neutron scattering (INS) measurements reveal quantum critical
TLL behavior of a coupled
$S=1/2$
chain compound YbAlO$_3$
with nearly isotropic (Heisenberg) intrachain exchange couplings~\cite{Wu_NC:2019}. A surprising observation is an incommensurate AFM state induced by the applied magnetic field. In this phase, the modulation of the ordering wave vector is proportional to the magnetization, which is a
characteristic of the LSDW order in coupled 
Ising anisotropic XXZ chains. This leads to the puzzle 
on the origin of the incommensurate AFM order in Heisenberg chains. A clue from both experimental observation and theoretical analysis is the relevance of the Ising anisotropic interchain coupling~\cite{Wu_NC:2019,Agrapidis_PRB:2019}. Still a generic open question is that
how the interchain Ising anisotropy would affect the phase diagram and low-energy excitations of Heisenberg chains. This poses a major challenge to existed theories based on the interchain mean-field approximation where the interchain fluctuations are neglected~\cite{Okunishi_PRB:2007}.

To tackle
these issues, in this letter we study the field-induced phase diagram of
$S=1/2$
Heisenberg chains with Ising anisotropic interchain couplings by
using large-scale quantum Monte Carlo (QMC) simulations.
Our results unambiguously show that the interchain interactions enhance
longitudinal spin correlations to stabilize an incommensurate LSDW.
With increasing field, the ground state transforms 
to a TAF state, and 
is fully polarized 
above 
a quantum critical point (QCP) controlled by the (3+2)D XY universality.
Increasing temperature from the QCP, the scaling of thermal energy and NMR relaxation rate demonstrate that the system experiences
a clear 
dimension crossover 
to the universal TLL behavior, 
exhibiting rich physics and fine
structure of the quantum criticality.
We then propose NMR measurements as 
a means to probe
the 
ground states
and 
related low-energy excitations in YbAlO$_3$ and other Q1D antiferromagnets.

{\it Model and method.~}
We consider a model defined on a three-dimensional (3D) cubic lattice for the
$S=1/2$
Heisenberg spin chains with weak interchain couplings of the XXZ type under a longitudinal ($z$-direction) magnetic field. The Hamiltonian reads as
\begin{eqnarray}\label{Eq:Ham}
H &=& {J_c}\sum\limits_{i} {{{\vec S}_i} \cdot {{\vec S}_{i+c}}} - g{\mu _B}H\sum\limits_i {S_i^z} \nonumber\\
&+& {J_{ab}}\sum\limits_{i,\delta=\{a,b\}}
\left[\varepsilon \left( {S_i^xS_{i+\delta}^x + S_i^yS_{i+\delta}^y} \right)
+ S_i^zS_{i+\delta}^z\right].
\end{eqnarray}
Here $\vec{S}_i=\{S^x_i,S^y_i,S^z_i\}$ is an $S=1/2$ spin operator defined on site $i$. $J_c$ and $J_{ab}$ are respectively the intrachain (along $c$ axis) and interchain exchange couplings between the nearest neighbor spins.
$\varepsilon$ is a parameter characterizing the spin anisotropy of the interchain coupling. $g$ is the gyromagnetic factor, ${\mu _B}$ is the Bohr magneton,
and $H$ is the applied magnetic field. We take $J_c$ as the energy unit and define the reduced temperature $t=T/J_c$ and reduced field $h=g\mu_B H/J_c$.
For Ising anisotropic interchain interaction, $\varepsilon<1$. Here we take $\varepsilon  = 0.25$ and $J_{ab} = 0.2 J_c$ for demonstration.
The effects of $\varepsilon$ and $J_{ab}$ on the phase diagram of the system will be discussed later.
To study the model in Eq.~\eqref{Eq:Ham} we perform numerically exact quantum Monte Carlo (QMC) simulations based on the stochastic series expansion (SSE) algorithm~\cite{Syljuasen_PRE:2002,Alet_PRE:2005}. In the simulations, the largest system size is $32 \times 32 \times 256$ and the lowest temperature accessed is $t=0.003$.

\begin{figure}[!t]
\includegraphics[
width=8.5cm]{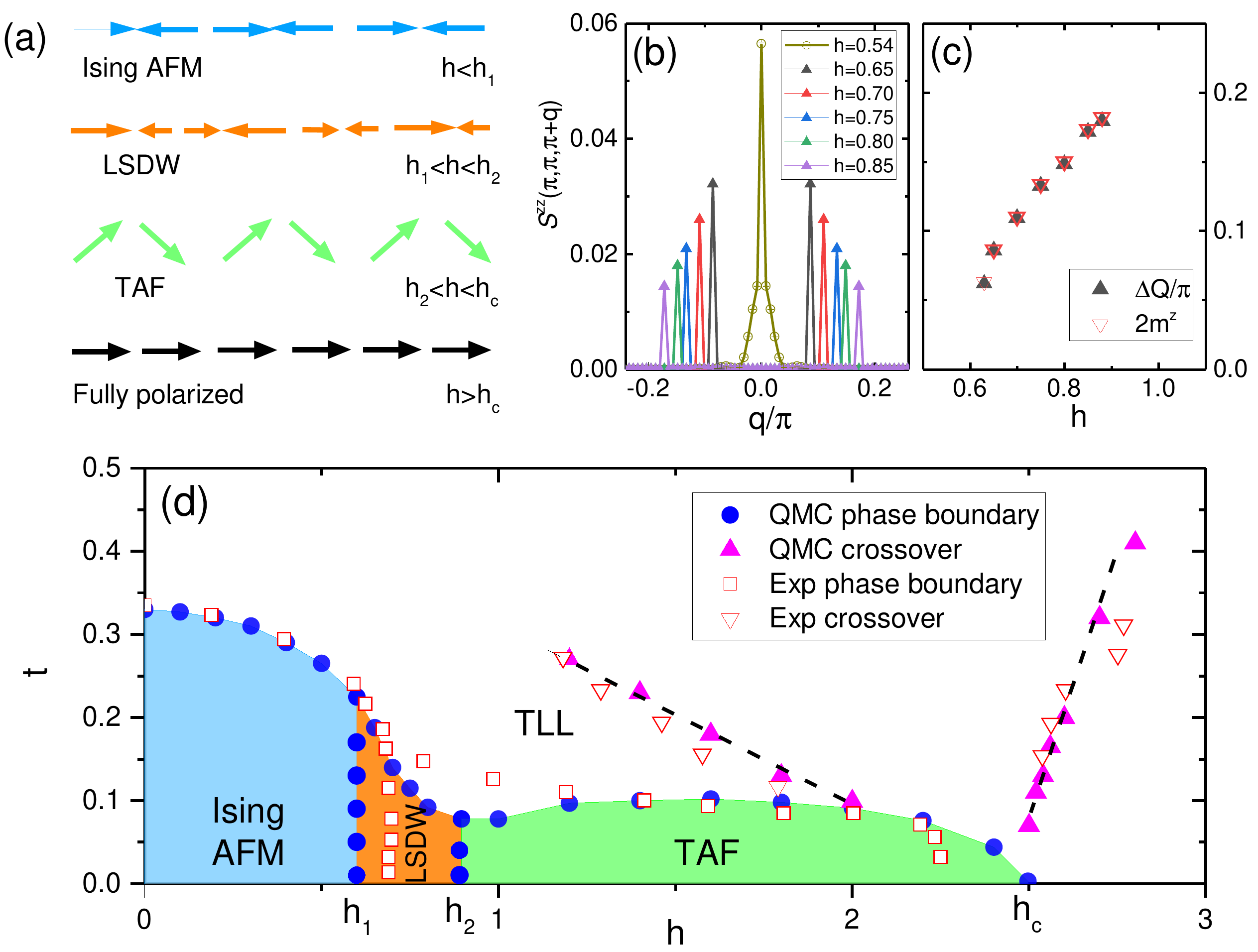} 
\caption{(a): 
Sketch of spin patterns along a chain in low-temperature phases. (b): Longitudinal spin structure factor at various fields. The splitting of the peak signals the LSDW order.
(c): Relation between the incommensurate ordering wave vector and the magnetization in the LSDW phase. (d): Thermal phase diagram of the model in Eq.~\eqref{Eq:Ham} obtained by QMC simulations. Filled circles denote the phase boundaries. The order-disorder transition at each field is determined by the peak position of the temperature dependent specific heat data (Fig.S1~\cite{SM}), while transitions between ordered phases are determined by the change of ordering wave vectors in calculated spin structure factors. Also shown are the adapted experimental phase boundary data (open squares) for YbAlO$_3$, taken from Ref.~\cite{Wu_NC:2019}. The filled and open triangles respectively show the calculated and adapted experimental crossover temperatures in the disordered regime close to the QCP at $h_c$. The dashed lines are linear fits.}
\label{Fig:1}
\end{figure}

{\it Phase diagram and the LSDW phase.~} Our main results are summarized in the phase diagram of Fig.~\ref{Fig:1}(d). At low temperatures, three ordered phases appear sequently with increasing field, and the spin patterns along a chain in these phases are illustrated in Fig.~\ref{Fig:1}(a). An Ising AFM phase with ordered moments aligned in the $z$ direction is stabilized for $h<h_1\approx0.6$, and a LSDW state with incommensurate longitudinal spin correlations is stabilized for field regime $h_1<h<h_2\approx0.89$. For $h>h_2$ the ground state becomes a TAF, which is a canted AFM state with staggered transverse spin correlations. Further increasing the field, the spins become fully polarized for $h>h_c\approx2.50$. The QPT at $h_c$ is continuous, while the transitions associated with the LSDW order at $h_1$ and $h_2$ are both first-order.

\begin{figure}[!t]
\includegraphics[
width=8.5cm]{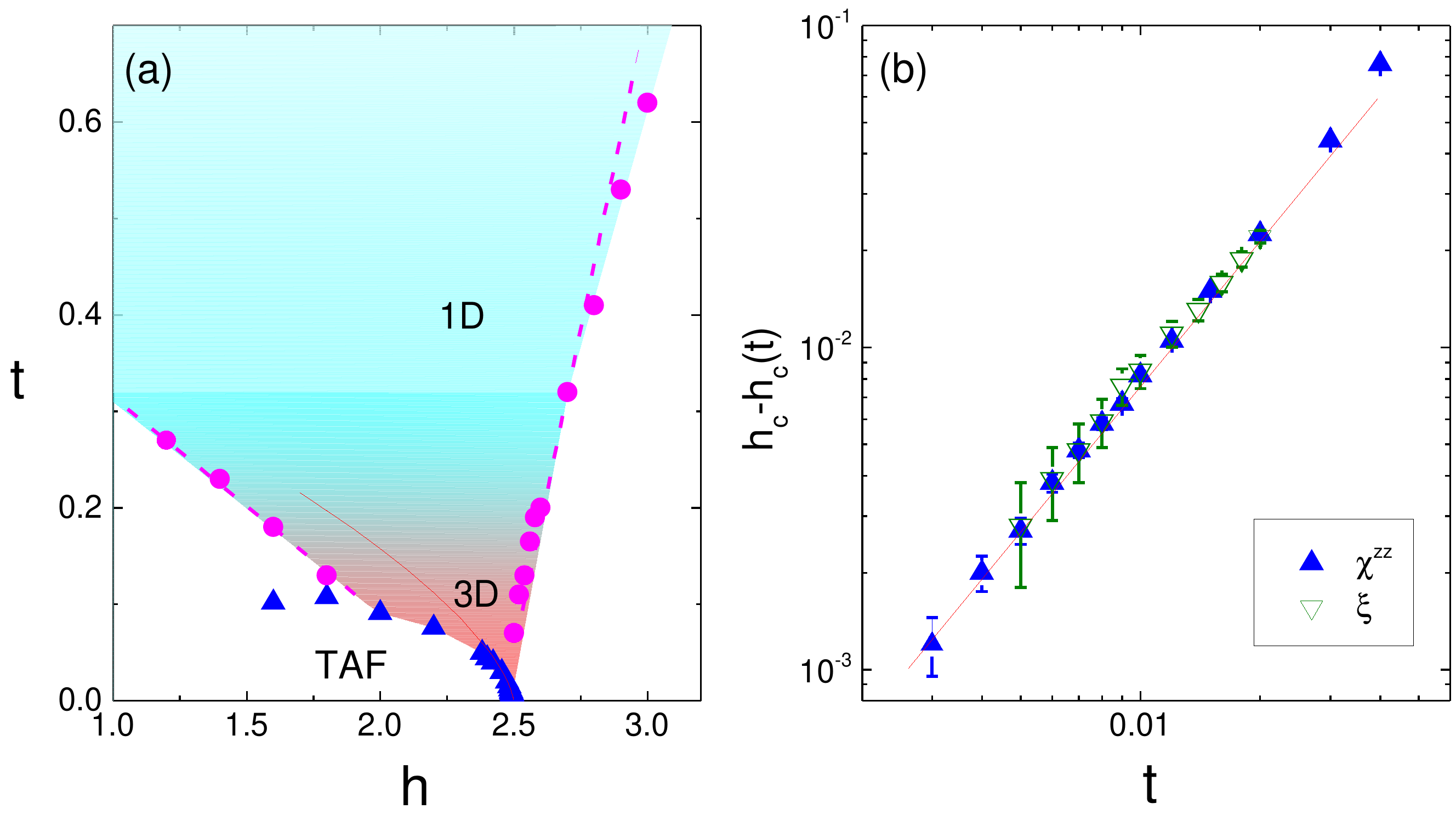} 
\caption{
(a): Phase boundary (blue triangles) and crossover (red circles) near the QCP. The red solid line is a power-law fit to $t\sim (h_c-h)^{2/3}$, and the dahsed lines are linear fits. The color scheme shows a 1D-3D crossover in the quantum critical regime. 
(b): Scaling of the critical fields $h_c(t)$ near the QCP, determined from susceptibility and correlation length data. The line is a fit $h_c-h_c(t)\sim t^{3/2}$.}
\label{fig:2}
\end{figure}

To examine the nature of the ordered states, we calculate the normalized longitudinal and transverse spin structure factors
\begin{eqnarray}
&& \mathcal{S}^{zz} (\mathbf{q}) = \frac{1}{N^2} \sum\limits_{ij} {e^{i\mathbf{q} \cdot ( \mathbf{r}_i - \mathbf{r}_j)}\left\langle S_i^z S_j^z \right\rangle }, \label{Eq:Szz} \\
&& \mathcal{S}^{xy} (\mathbf{q}) = \frac{1}{2N^2} \sum\limits_{ij} {e^{i\mathbf{q} \cdot ( \mathbf{r}_i - \mathbf{r}_j)}\left\langle S_i^x S_j^x + S_i^y S_j^y \right\rangle }. \label{Eq:Sxy}
\end{eqnarray}
The Ising AFM order is signaled by a peak of $\mathcal{S}^{zz}(\mathbf{q})$ at $\mathbf{q}=(\pi,\pi,\pi)$. When $h>h_1$ we find that the peak splits into two located at incommensurate $\mathbf{q}=(\pi,\pi,\pi\pm\Delta Q)$ (Fig.~\ref{Fig:1}(b)). In this incommensurate AFM phase the ordering wave vector varies with increasing field, satisfying $|\Delta Q | = 2\pi\langle {m^z} \rangle$ (Fig.~\ref{Fig:1}(c)), a characteristic reflecting the Q1D TLL physics of the LSDW state~\cite{Okunishi_PRB:2007}. 
This confirms that the incommensurate order is indeed a LSDW, which in this model arises 
from the enhancement of longitudinal correlations by interchain Ising anisotropy. For $h>h_2$, the peak of $\mathcal{S}^{zz}$ is suppressed and the ground state changes to the TAF with a peak of $\mathcal{S}^{xy}(\mathbf{q})$ at $\mathbf{q}=(\pi,\pi,\pi)$, as shown in Fig.S2 of Supplemental Material (SM)~\cite{SM}.

\begin{figure}[!t]
\includegraphics[
width=8.5cm]{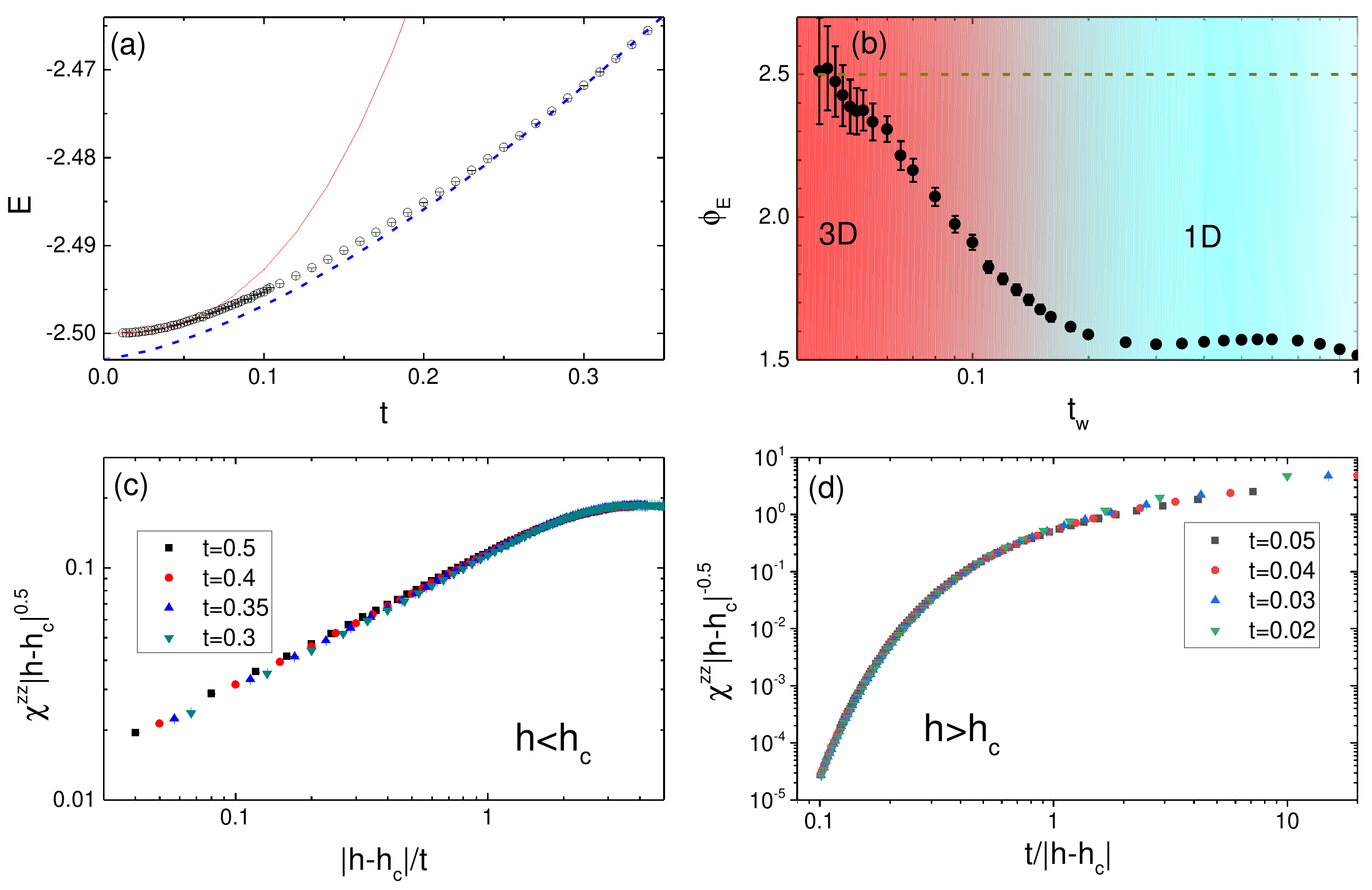} 
\caption{(a): Temperature evolution of the thermal energy at $h_c$. The solid and dahsed lines are power-law fits $E\sim t^{\phi_E}$ with $\phi_E=5/2$ and $\phi_E=3/2$, respectively. (b): Windowing estimate on $\phi_E$ showing a 1D-3D crossover. (c): Scaling of susceptibility showing effective 1D quantum critical TLL behavior at high temperatures using $h<h_c$ data. (d): Scaling of susceptibility showing the $(3+2)$D nature of the QCP at low temperatures using $h>h_c$ data.}
\label{fig:3}
\end{figure}
{\it Quantum criticality.} The QPT at $h_c$ takes place when the TAF order is suppressed. Since the TAF order breaks the spin $U(1)$ symmetry, the transition can be viewed as a magnetic BEC~\cite{Giamarchi_NP:2008,Zapf_RMP:2014} with a dynamical exponent $z=2$. The QCP then belongs to the $(3+2)$D XY universality class. To check this, we first study the scaling behavior of critical field $h_c(t)$ at low temperatures. As shown in Fig.~\ref{fig:2}, $h_c(t)$ data determined from either the peak of field dependent susceptibility $\chi^{zz}(h)=\partial m^z/\partial h$ or the correlation length $\xi\sim L$ (Fig.S3 and Fig.S4~\cite{SM}) follow the scaling relation of $3d$ BEC, $h_c-h_c(t)\sim t^{d/2}=t^{3/2}$. We then study the finite-temperature crossover in the vicinity of the QCP. At each field the temperature dependent susceptibility $\chi^{zz}(t)$ develops a broad peak, and the peak position defines the crossover temperature $T_{cr}$ (Fig.S4~\cite{SM}). Near a QCP, $T_{cr}\sim |h-h_c|^{\nu z}$, where $\nu$ is the correlation length exponent. From Fig.~\ref{fig:2}(a) we find that $T_{cr}\sim |h-h_c|$ on both sides of the QCP, consistent with the $(3+2)$D XY universality $z=2$ and $\nu=1/2$.

Owing to its Q1D structure, the system exhibits finite temperature 1D-3D crossover. This is clearly shown in the scaling of thermal energy, $E=\langle H \rangle/N$, right at the critical field $h_c$. Since $dE/dt=C\sim t^{d/z}$, where $C$ is the specific heat, we expect $E\sim t^{\phi_E}$ with $\phi_E=d/z+1=5/2$. But as shown in Fig.~\ref{fig:3}(a), this scaling fits only at low temperatures for $t\lesssim 0.06$. And for $t\gtrsim 0.2$, $\phi_E\approx3/2$, implying an effective dimension $d_{eff}=1$. Careful windowing analysis~\cite{Sebastian_PRB:2005} in Fig.~\ref{fig:3}(b) finds a gradual increase of $\phi_E$ from about $3/2$ to $5/2$ for $0.06\lesssim t \lesssim 0.2$, clearly indicating a 1D-3D crossover in this temperature regime. The 1D-3D crossover gives rise to rich quantum scaling behaviors. For example, the genuine 3D nature of the QCP is inherent in the low-temperature scaling of susceptibility data in the disordered phase (Fig.~\ref{fig:3}(d)), which satisfies
 $\chi^{zz} \sim |h-h_c|^{\nu (d+z) -2} X \left( \frac{t}{|h-h_c|^{\nu z}}\right)$ with $d=3$, $\nu=1/2$, and $z=2$.
In the quantum critical regime it is expected that 
 $\chi^{zz} \sim |h-h_c|^{d/z+1-2/\nu z} \tilde{X} \left( \frac{|h-h_c|}{t^{1/\nu z}}\right)$ with the same exponents at low temperatures.
But the scaling of QMC data above the dimension crossover temperature in Fig.~\ref{fig:3}(c) are consistent with $d_{eff}=1$, $\nu=1/2$, and $z=2$, characterizing a quantum critical TLL behavior.

{\it NMR relaxation rate $1/T_1$ and the TLL behavior.}
In the TLL regime of an XXZ chain the spin correlations decay algebraically as
$ \langle S_0^z S_r^z \rangle  - (m^z)^2 \sim \cos ( 2 k_F r ) r^{-1/\eta} $ 
and
$ \langle S_0^x S_r^x \rangle \sim (-1)^r r^{-\eta} $,
where $k_F = \pi (1/2-m^z)$, denoting the Fermi wave number of pseudo-fermions mapped from the spin model by a Jordan-Wigner transformation, and 
the Luttinger exponent $\eta$ determines the decay rate. For a Heisenberg chain, $\eta<1$ for all fields and the staggered transverse fluctuations are always dominant. On the other hand, for an Ising anisotropic XXZ chain, there is an $\eta$ inversion at field $h_{inv}$, \emph{e.g.} $\eta>1$ for $h<h_{inv}$ where longitudinal fluctuations dominate and $\eta<1$ for $h>h_{inv}$ where the dominant fluctuations turn to transverse ones~\cite{Okunishi_PRB:2007}.

\begin{figure}[!t]
\includegraphics[
width=8.5cm]{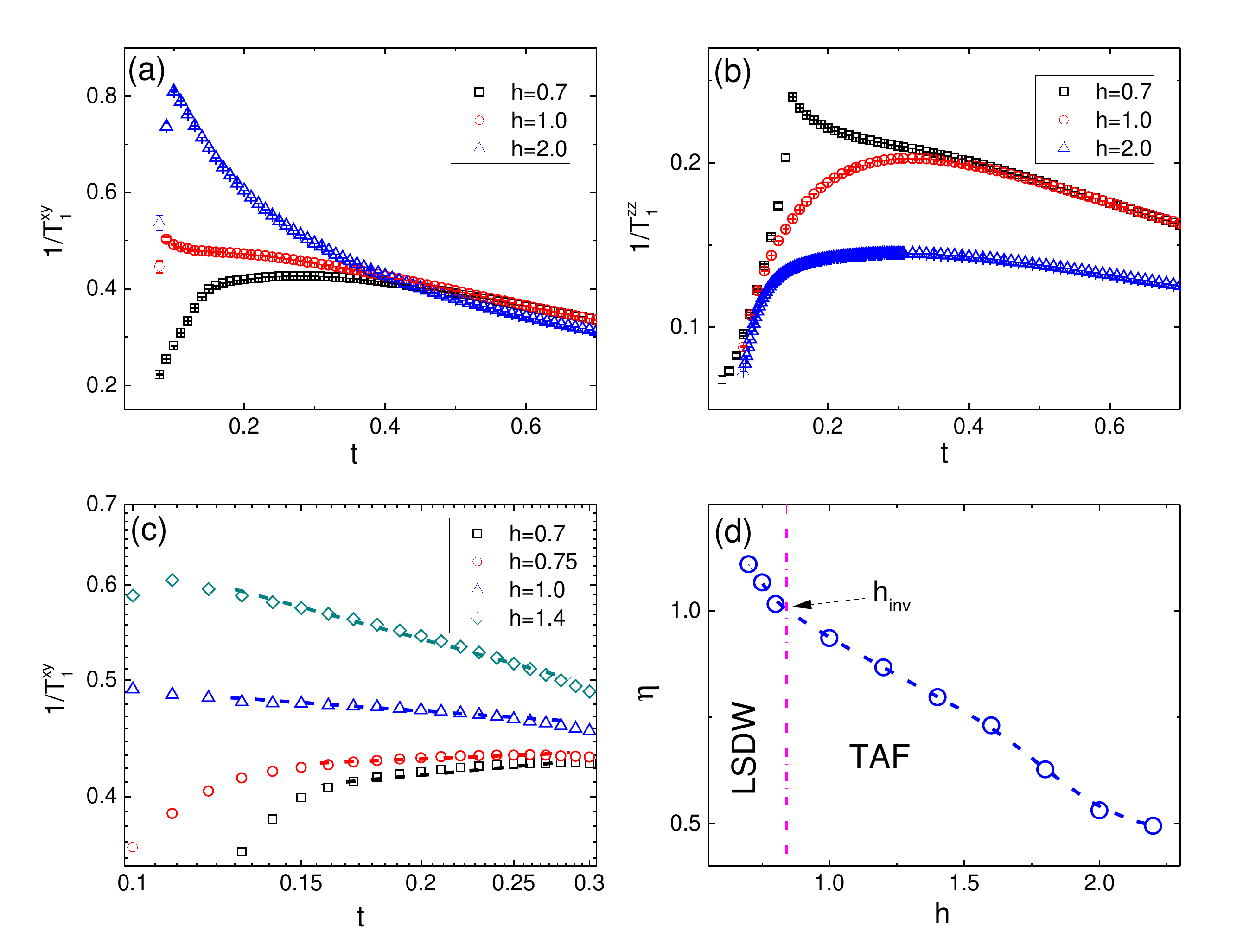} 
\caption{(a): Temperature dependence of the calculated transverse part of the NMR relaxation rate, $1/T_1^{xy}$, at various fields. (b): Longitudinal part $1/T_1^{zz}$. (c): Same as (a) but in double-logarithmic scale, showing TLL behavior above the ordering temperature. The lines are power-law fits. (d): Extracted $\eta$ exponent from $1/T_1^{xy}$ data. $h_{inv}$ is the crossover field where the dominant fluctuation changes from longitudinal ($\eta>1$) to transverse ($\eta<1$) with increasing field. It is found $h_{inv}\approx h_2$, which separates the LSDW and TAF ground states.}
\label{fig:4}
\end{figure}

To examine whether the TLL behavior near $h_c$ extends to lower fields and to determine the dominant spin fluctuations associated with magnetic ordering we calculate the NMR spin-lattice relaxation rate $1/T_1$, which probes the low-energy spin fluctuations of a magnetic system~\cite{
Coira_PRB:2016,Dupont_PRB:2016}. For simplicity we set the nuclear gyromagnetic ratio and the hyperfine coupling to be unity, and define the longitudinal ($zz$) and transverse ($xy$) relaxation rates as
 $1/T_1^{\alpha\alpha} = \int d\mathbf{q} \Im\chi^{\alpha\alpha}(\mathbf{q},\omega_0)/\beta\omega_0$,
where $\beta=1/t$, $\Im\chi^{\alpha\alpha}(\mathbf{q},\omega_0)$ is the imaginary part of the dynamical susceptibility at NMR frequency $\omega_0\rightarrow0$, $\alpha=x,y,z$, and $1/T_1^{xy}=1/T_1^{xx}+1/T_1^{yy}$. To avoid handling the analytical continuation in QMC simulations, we further adopt an approximation~\cite{Randeria_PRL:1992,Roscilde},
\begin{equation}\label{Eq:T1QMC}
1/T_1^{\alpha\alpha}\approx\frac{2}{\pi t}\sum_i \langle \delta S^\alpha_i(\beta/2) \delta S^\alpha_i(0)\rangle,
\end{equation}
where $\delta S^\alpha_i=S^\alpha_i-\langle S^\alpha_i \rangle$. As shown in Fig.S5~\cite{SM}, this approximation gives reasonable results for a Heisenberg chain.

The results for the 3D model are shown in Fig.~\ref{fig:4}(a) and (b). At $h=0.7$ where the ground state is a LSDW, the temperature dependent $1/T_1^{zz}$ develops a prominent peak at the ordering temperature, signaling enhanced critical fluctuations. But $1/T_1^{xy}$ only shows a kink at the transition and is significantly suppressed in the ordered phase. At higher fields where the ground state is the TAF, the peak feature is seen in $1/T_1^{xy}$, and $1/T_1^{zz}$ drops rapidly in the ordered phase. Above the transition we find an algebraic temperature dependence of $1/T_1^{xy}$ (Fig.~\ref{fig:4}(c)), signaling a TLL behavior. In a TLL $1/T_1^{xy}\sim T^{\eta-1}$ according to bosonization results~\cite{Bouillot_Thesis:2011}. Fitting to this function, we extract the $\eta$ parameter at each field, as shown in Fig.~\ref{fig:4}(d). Surprisingly $\eta>1$ for $h\lesssim0.85$, indicating dominant longitudinal fluctuations in Heisenberg chains. With increasing field, $\eta$ decreases monotonically, and an $\eta$ inversion takes place at $h_{inv}\approx 0.85$. Interestingly, $h_{inv}\approx h_2$, which separates the LSDW and TAF ground states. The $\eta$ inversion and the peak feature of $1/T_1$ at transition indicate that the condensation of the dominant fluctuations leads to the corresponding type of magnetic order in this system.

{\it Discussions and Conclusion.~} Our QMC results provide the first clear evidence of a LSDW phase in a 3D model. 
The calculated $1/T_1$ data unambiguously show that this phase is stabilized by the enhanced incommensurate longitudinal fluctuations in Heisenberg chains. The enhancement of the longitudinal fluctuations originates from the interchain Ising anisotropy and this effect is beyond the conventional mean-field scenario~\cite{Okunishi_PRB:2007} in which the dynamical effects of the interchain couplings are ignored so that the transverse fluctuations always dominate.

When $h>h_{inv}$ the dominant spin fluctuations become transverse
and the TAF ground state is correspondingly stabilized. The transverse fluctuations also govern in the quantum critical regime where the system shows quantum critical TLL behavior at intermediate temperatures and converges to the $(3+2)$D XY universality at low temperatures. Such a scenario of quantum criticality generally holds for a broad class of weakly coupled XXZ spin chain systems that have the same symmetry as the model in Eq.~\eqref{Eq:Ham}.

For the model studied here, we find that the phase diagram is sensitive to the interchain coupling parameters $\varepsilon$ and $J_{ab}$. It is known that increasing $J_{ab}$ favors a TAF order owing to the spin-flop mechanism~\cite{Fisher_PRL:1974}. For fixed $J_{ab}$, the enhancement of longitudinal correlations and hence the stabilization of LSDW only take place when $\varepsilon$ is less than a critical value. As illustrated in Fig.S6~\cite{SM} the LSDW is
absent
for $\varepsilon\gtrsim0.5$ at $J_{ab}/J_c=0.2$. When $\varepsilon\rightarrow0$, however, the transverse correlations are only within each chain. Because the TAF order breaks a continuous $U(1)$ symmetry, it can not be stabilized in this limit. In this case the QPT is from the LSDW to the fully polarized phase and belongs to the $(3+1)D$ universality. But for a finite $\varepsilon$ we always find a TAF phase before the magnetization is fully saturated (see Fig.S6~\cite{SM}). Hence the LSDW is irrelevant to the QPT, which is always controlled by the $(3+2)$D XY universality.

In what follows we discuss implications of our results to YbAlO$_3$ and other related Q1D quantum magnets. It is found that the intrachain exchange coupling of YbAlO$_3$ is almost isotropic but the interchain one is dominant by dipole-dipole interaction containing strong Ising anisotropy~\cite{Wu_NC:2019,Wu_arXiv:2019}. This is fully captured by the Hamiltonian in Eq.~\eqref{Eq:Ham}, where the interchain Ising anisotropy is ensured by the finite $\varepsilon<1$. Taking the measured values $J_c\sim0.22$ meV
and $g\sim7.6$ \cite{Wu_NC:2019}, we can compare our results with experimental ones for YbAlO$_3$. 
As shown in Fig.~\ref{Fig:1}(d), the phase boundary of the model agrees qualitatively with the adapted experimental one. In particular, the LSDW state in our model naturally explains the observed unusual incommensurate AFM order. Note that INS measurement suggests a ferromagnetic interchain coupling~\cite{Wu_NC:2019}. 
Though for demonstration we take $J_{ab}>0$, the stabilization of the LSDW phase and the agreement on the phase boundary indicate that our model has already captured the essential physics of the system, and the results for $J_{ab}<0$ are qualitative the same.
For the quantum criticality, the linear scaling of $T_{cr}$ and the quantum critical TLL behavior obtained in our theory are also observed in YbAlO$_3$~\cite{Wu_NC:2019}, while the predicted genuine $(3+2)$D XY universality, which also well applied to other coupled XXZ spin chains such as (Ba,Sr)V$_2$Co$_2$O$_8$, can be tested by future experiments.

Our theory also predicts a TAF order for $h>h_2$. In real materials, the LSDW and TAF orders may coexist or are phase separated~\cite{Grenier_PRB:2015, Bera_SCVO:2019}.
This possibility and the strong anisotropic gyromagnetic tensor in YbAlO$_3$ complicates the detection of transverse spin correlations and the related TAF order by INS measurements~\cite{Wu_NC:2019}. In light of the theoretical results, we hereby propose to probe the dominant low-energy spin fluctuations and associated magnetic ordering by measuring NMR $1/T_1$. Even when the system is highly anisotropic such that only $1/T_1^{zz}$ or $1/T_1^{xy}$ is detectable, the peak or kink feature in the temperature dependent $1/T_1$ near the ordering temperature can still tell the dominant fluctuations and the associated magnetic order, as shown in Figs.~\ref{fig:4}(a) and (b). Moreover, it would be interesting to examine the possible $\eta$ inversion by a careful study on the $1/T_1$ data above the ordering temperature. Such a study can also provide important information on the dominant fluctuations and the underlying magnetic ground state. Given the universal property of TLL, similar analysis can be applied on other related Q1D systems, such as (Ba,Sr)V$_2$Co$_2$O$_8$, where the dominant fluctuations and associated long-range magnetic orders are still under debate~\cite{Klanjsek_PRB:2015,Grenier_PRB:2015}.

In conclusion, we study the field-induced phase diagram and quantum criticality of $S=1/2$ AFM Heisenberg chains with Ising anisotropic interchain couplings. We find the interchain Ising anisotropy enhances incommensurate AFM correlations, stabilizing a LSDW ground state at low fields. The transverse spin correlations dominate at high fields and a TAF ground state is stabilized. This leads to a QCP controlled by a $(3+2)$D XY universality but displaying a finite-temperature 1D-3D crossover. The calculated NMR relaxation rates show enhanced critical fluctuations at magnetic ordering, and the enhancement takes place at particular channel relevant to the underlying magnetic order. Above the ordering temperature, the system exhibits universal TLL behavior and shows an $\eta$ inversion with increasing field, where the dominant spin fluctuation changes from longitudinal to transverse. These features make NMR an ideal probe for the spin fluctuations and associated ground states of coupled spin chains. Our findings thus shed light on future experimental and theoretical studies on YbAlO$_3$ and other Q1D quantum magnets.

{\it Acknowledgments.} We thank useful discussion with L. S. Wu. This work was supported by the Ministry of Science and Technology of China (Grant No. 2016YFA0300504), the National Natural Science Foundation of China (Grants No. 11674392, and 51872328), the Fundamental Research Funds for the Central Universities, the Research Funds of Renmin University of China (Grant No. 18XNLG24), the Science and Technology Commission of Shanghai Municipality Grant No. 16DZ226020, and the Outstanding Innovative Talents Cultivation Funded Programs of Renmin University of China. R.Y. acknowledges hospitality at ENS de Lyon, France. J.W. acknowledges additional support from a Shanghai talent program.

\clearpage
\setcounter{figure}{0}
\makeatletter
\renewcommand{\thefigure}{S\@arabic\c@figure}
\onecolumngrid
\section*{ SUPPLEMENTAL MATERIAL -- Phase diagram and 
quantum criticality of Heisenberg spin chains with Ising-like interchain couplings -- Implication to YbAlO$_3$}


\begin{figure}[h!]
\centering
\includegraphics[
width=12.0cm]{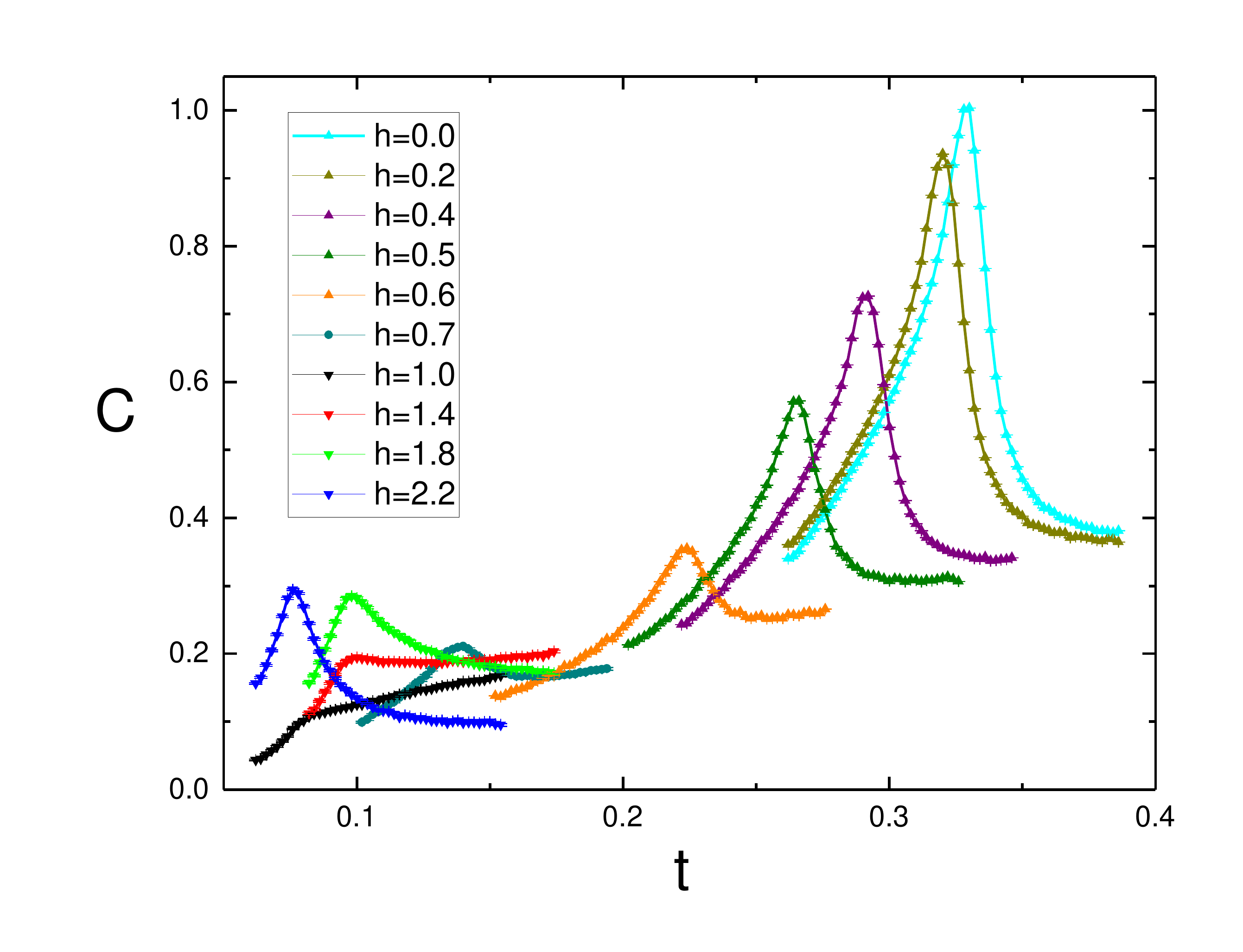}
\caption{(Color online) Temperature dependence of specific heat $C$ at various field values, which is used to determine the phase boundary in Fig.~1 of the main text. At each field, the transition to an AFM state is signaled as either a peak or a kink feature.
}
\label{Sfig:1}
\end{figure}
\begin{figure}[h!]
\centering
\includegraphics[
width=12.0cm]{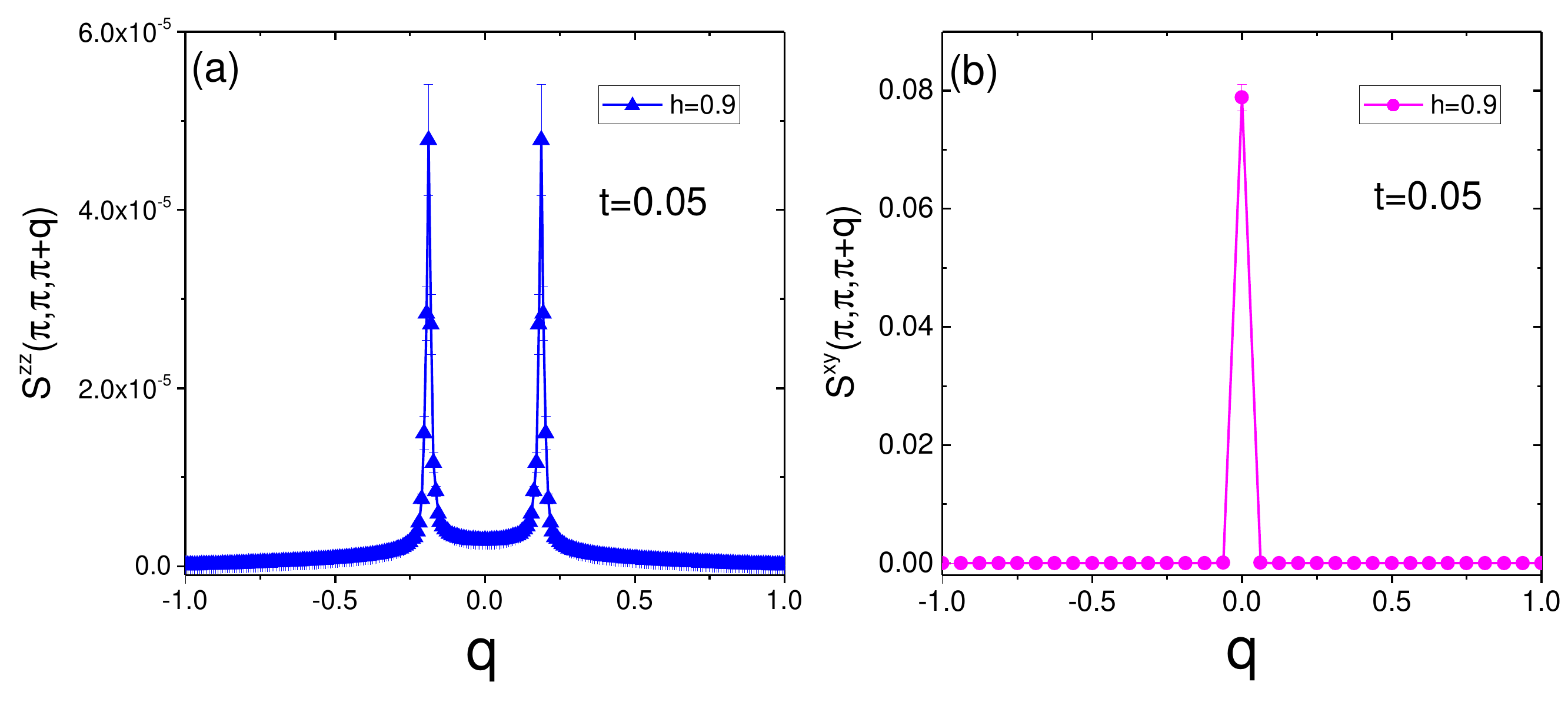}
\caption{(Color online) $q$ dependence of the longitudinal [in (a)] and transverse [in (b)] spin structure factors, $\mathcal{S}^{zz}(\pi,\pi,\pi+q)$ and $\mathcal{S}^{xy}(\pi,\pi,\pi+q)$, respectively, at $t=0.05$ and $h=0.9$ in the TAF phase.}
\label{Sfig:2}
\end{figure}

\begin{figure}[h!]
\centering
\includegraphics[
width=12.0cm]{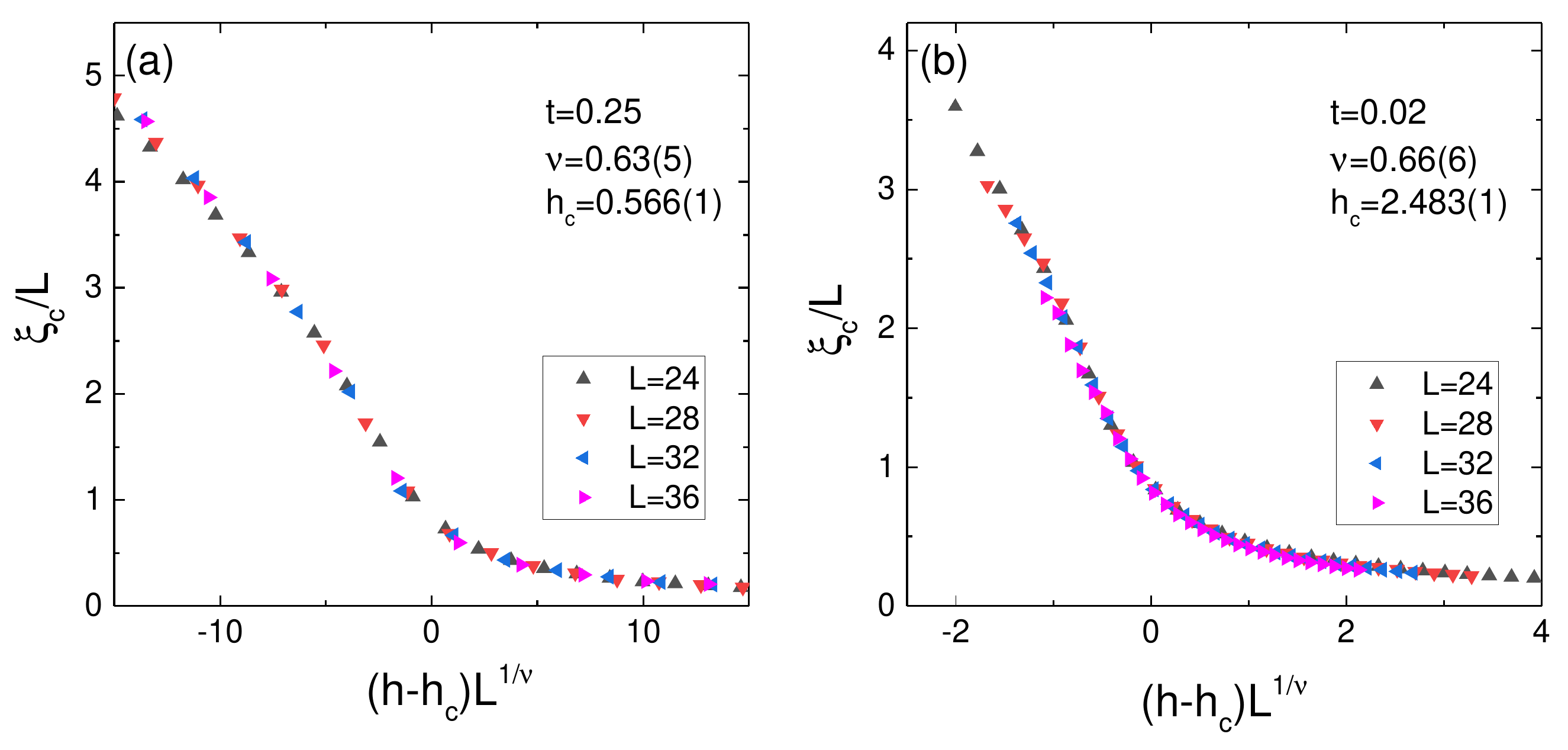}
\caption{(Color online) Finite-size scaling of the correlation length along the $c$ axis, $\xi_c$, at transition to the TAF phase. The determined critical field $h_c$ is plotted in Fig.~3(b) of the main text, and the extracted correlation length exponent $\nu$ agrees with the value of 3D XY universality within error bar.}
\label{Sfig:3}
\end{figure}

\begin{figure}[h!]
\centering
\includegraphics[
width=12.0cm]{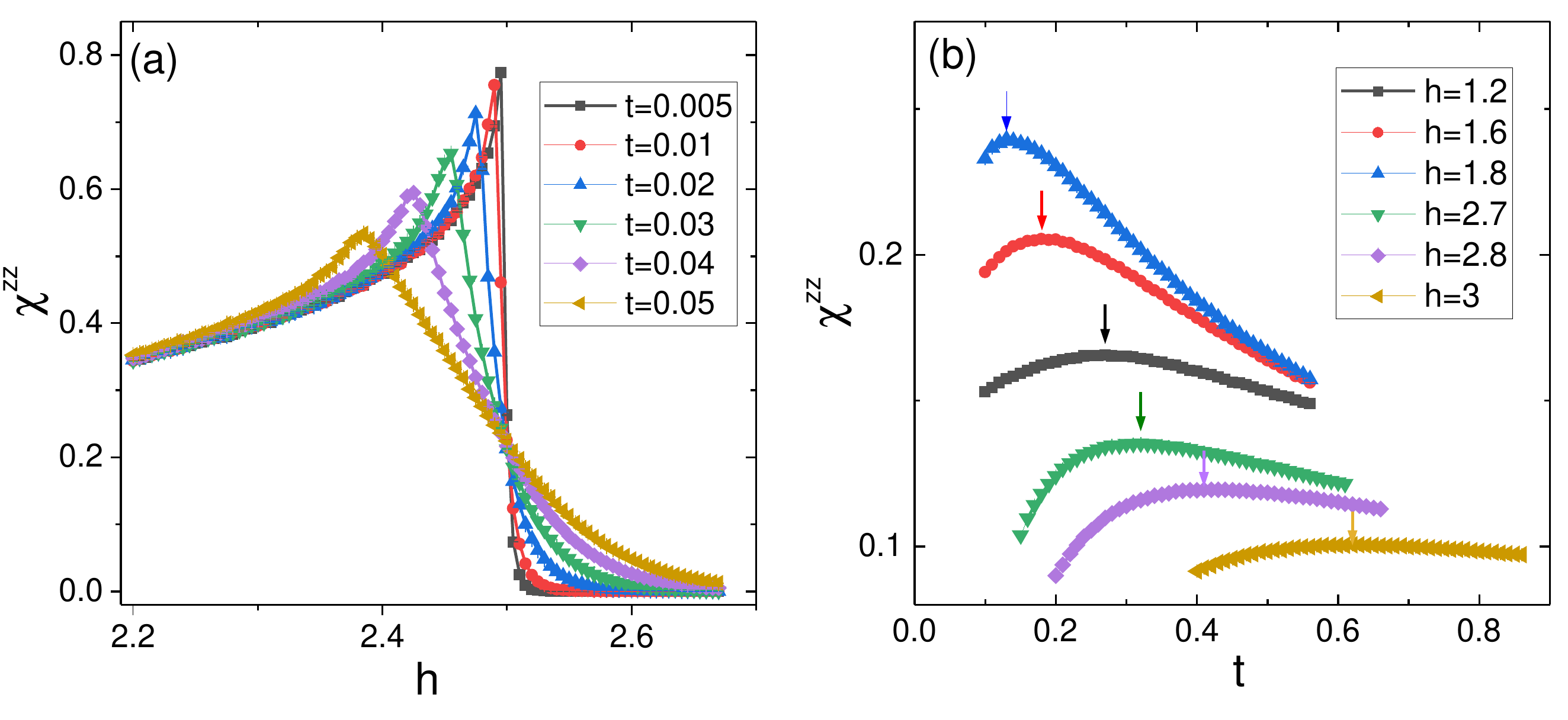}
\caption{(Color online) (a): Field dependence of susceptibility $\chi^{zz}$ at low temperatures, where the peak determines the critical field $h_c(t)$. (b): Temperature dependence of $\chi^{zz}$ above the ordering temperatures. The peak position (pointed by an arrow) determines the crossover temperature $T_{cr}$ in Fig.~1(d) and Fig~2(a) of the main text.}
\label{Sfig:4}
\end{figure}

\begin{figure}[!h]
\centering
\includegraphics[
width=12.0cm]{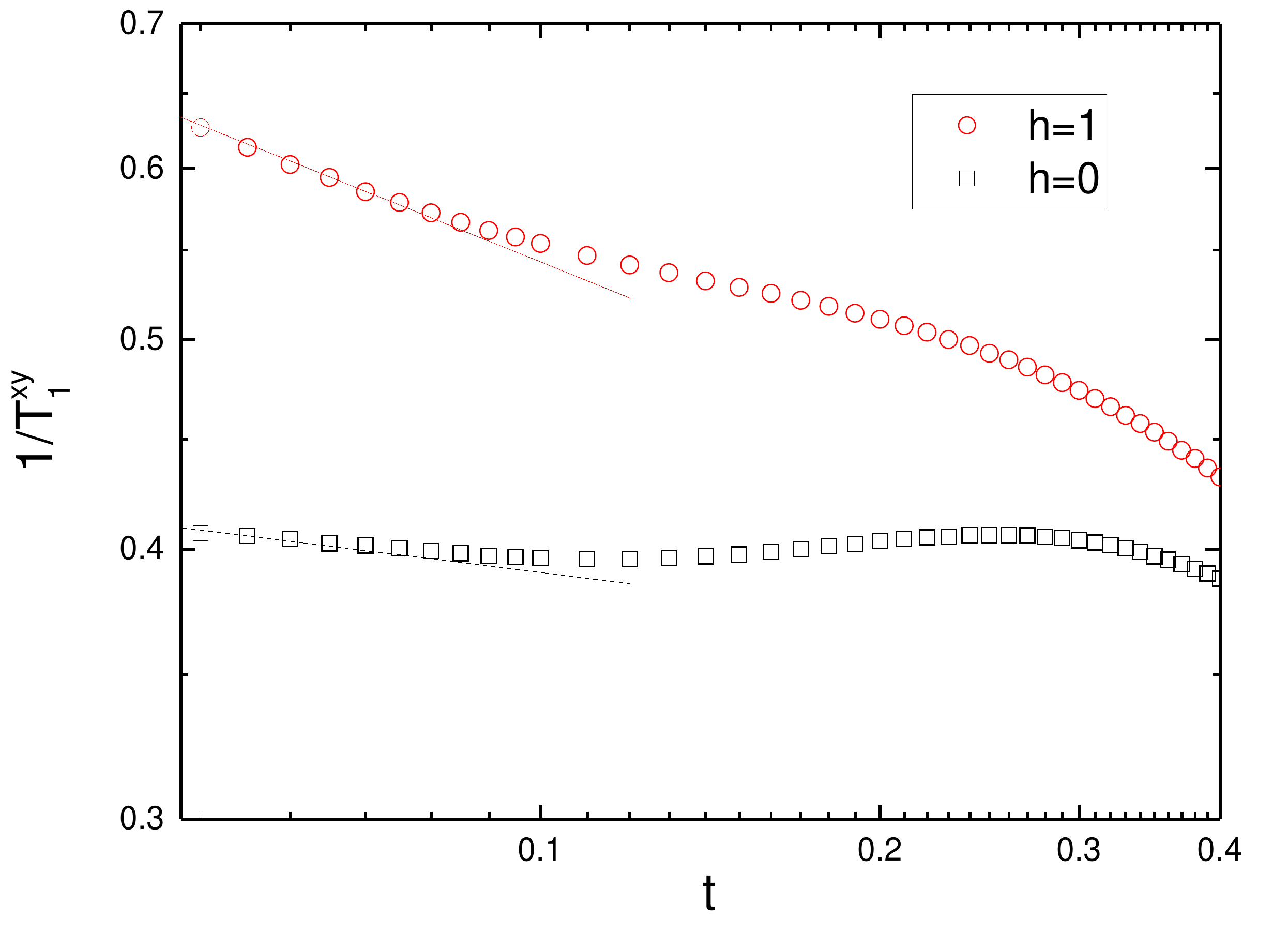}
\caption{(Color online) Temperature dependence of the calculated transverse part of the NMR relaxation rate, $1/T_1^{xy}$, for a single Heisenberg chain at $h=0$ and $h=1$, respectively. The lines are power-law fits from analytical results in Refs.~\cite{Coira_PRB:2016,Dupont_PRB:2016}.}
\label{Sfig:5}
\end{figure}

\begin{figure}[!h]
\centering
\includegraphics[
width=12.0cm]{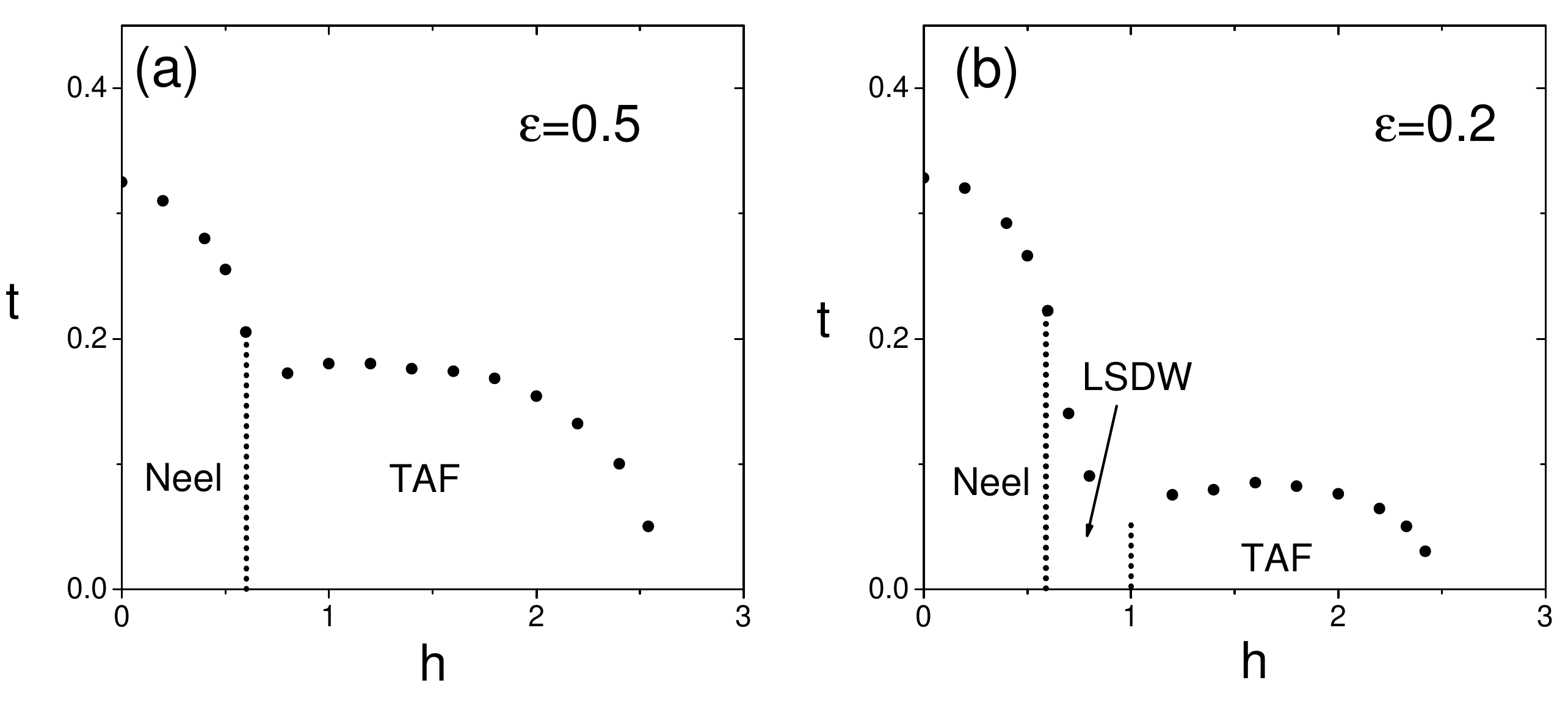}
\caption{(Color online) Thermal phase diagram of the model in Eq.~(1) of the main text for $J_{ab}/J_c=0.2$ and $\varepsilon=0.5$ [in (a)] and $\varepsilon=0.2$ [in (b)]. No LSDW phase is stabilized at $\varepsilon=0.5$. The TAF phase is stabilized in both cases.}
\label{Sfig:6}
\end{figure}


\end{document}